\documentstyle[12pt,fullpage,epic,eepic,fullpage,float,psfig]{article}
\restylefloat{figure}
\restylefloat{table}

\title{Numerical Simulation of Grain Boundary Grooving By Level Set
Method}
\author{M. Khenner$^{1}$ ~~A. Averbuch$^{1}$~~ M. Israeli$^{3}$~~M. Nathan$^{3}$\\
$^{1}$Department of Computer Science \\
School of Mathematical Sciences \\
Tel Aviv University, Tel Aviv 69978, Israel\\~\\
$^{2}$ Faculty of Computer Science \\
Technion, Haifa 32000, Israel \\~\\
$^{3}$Department of Electrical Engineering-Physical Electronics \\
Faculty of Engineering \\
Tel Aviv University, Tel Aviv 69978, Israel}
\date{}

\newcommand{\Section}[1]{\setcounter{equation}{0} \section{#1}}
\newcommand{\rf}[1]{(\ref{#1})}
\newcommand{\beq}[1]{ \begin{equation}\label{#1} }
\newcommand{\eeq}{\end{equation} }

\begin{document}
\pagestyle{plain}
\maketitle
\date{}
\begin{center}
{\em Submitted to Journal of Computational Physics}
\end{center}
\abstract{A numerical investigation of grain-boundary grooving
by means of a Level Set method is carried out.
An idealized
polygranular interconnect
which consists of grains separated by parallel grain boundaries
aligned normal to the average orientation of the surface
is considered.
The surface diffusion is the only physical mechanism assumed.
The surface diffusion is driven by surface curvature gradients, and
a fixed surface slope and zero atomic flux are assumed at the groove
root.
The corresponding mathematical system is an initial boundary
value problem for a two-dimensional Hamilton-Jacobi
type equation.
The results obtained are in good agreement with both
Mullins' analytical ``small slope" solution of the linearized problem
\cite{MULLINS57}
(for the case of an isolated grain boundary) and with solution for
the periodic array of grain boundaries (Hackney \cite{HACKNEY}).}

\Section{Introduction}
This paper presents the results of
our work on numerical modeling and simulation of
grain-boundary (GB) grooving by surface diffusion.
Our ultimate goal is to develop and test a fast numerical approach
for
the simulation of formation and propagation of groove-like
defects in thin film interconnects
used in microelectronics (ME).

In modern ME industry, the
quality and reliability of ME integrated circuits have become no less
important than their performance. Some of the most vulnerable elements
of
ME circuits, susceptible to several types of mechanical failures,
are the interconnects.
These are metallic conductors which connect the active elements.
%They may occupy about $3/4$ of a chip area
%and take about half the cycle time in the faster circuits used in
%computers.

The defects (due to the small cross-section, high current density, 
mechanical stresses and presence of GBs acting as fast diffusion pathways) lead
to the loss of electrical and mechanical integrity, i.e. to line opens or shorts.
Thus, such defects are one of the main
reliability concerns in advanced integrated circuits.

\subsection{Mechanisms of Mechanical Failure in Interconnect Lines}

In this section we describe some basic failure mechanisms in interconnects
and outline an appropriate physical model.

Many properties of polycrystalline materials are affected by the
intersection of GBs
with external surfaces, especially in the presence of applied or
internal
fields. Common examples are growth of GB grooves and cavities
%\cite{CHUANG-RICE,CKRS,KGFMB1,KGFMB2},
\cite{KGFMB1,KGFMB2},
stress voiding \cite{YOST} and electromigration
\cite{AIR,LZG,OHRING,ROSENBERG-OHRING}.

In the absence of an external potential field,
the GB atomic flux $I_{GB}=0$ and the corresponding groove
profile evolves via surface diffusion
under well-known
conditions of scale and temperature (the so-called Mullins' problem
\cite{MULLINS57}).
Mass transport by surface diffusion is driven by the surface Laplacian
of curvature.
Essentially, for convex surfaces, matter flows from high-curvature
regions,
while for concave surfaces the flow is from low curvature regions.
In order to solve surface-diffusion problems, four different approaches
have
been taken. We refer the interested reader to the article by Zhang and
Schneibel \cite{ZHANG-SCHNEIBEL}, where these
approaches are discussed and to the references
therein.

The physical origins of a GB flux may be gradients of the normal
stress at grain boundaries
%\cite{CHUANG-RICE,CKRS,GMW}
\cite{GMW}
and/or electromigration forces \cite{BLECH-HERRING}.
%The former effect is linear in the stress and hence should dominate the
%surface
%instability which depends quadratically on stress \cite{ATGS}.
GB grooving with a GB flux  in real thin film interconnects is 
a complex problem. It  requires sophisticated 
numerical modeling technique which can manage 
with such issues  as aperiodic arrays of GBs, anisotropy of 
the surface tension, GB migration,  formation  
of slits  with a local steady-state shape in the near-tip 
region  and bridging across the slits near their intersections  
with the surface left behind \cite{OHRING}.
Level Set Method seems to be a good candidate for  
addressing the problems, however it has never been used yet to 
this aim.  As the first step in application of LS Method  to the problem  
of grooving with EM flux, we test in this paper LS Method over two 
simple -and already solved-  grooving problems and compare the 
LS Method' results with those obtained previously in \cite{MULLINS57, HACKNEY}. 
First is classical Mullins' problem (GB grooving controlled by surface 
diffusion in an infinite bicrystal with a stationary  GB). 
Second is GB grooving by surface diffusion in the periodic GB array of 
stationary GBs. The electromigration flux will be taken 
into account in the next publication.  

%In the following,
%we consider an idealized polygranular
%interconnect
%which consists of grains separated by parallel GBs
%(GBs are aligned normal to the average orientation of the surface).
%We first focus on a simplest case of only two grains separated by GB.
%The periodic array of grains is studied in the end of Section 3.
%Electromigration
%forces are not taken into consideration in this paper. The
%treatment of electromigration is currently
%under development.

%In the following,
%we consider an idealized
%polygranular interconnect
%which consists of grains separated by parallel GBs. Electromigration
%forces are not taken into account.
%Our goal is {\bf a)} to obtain the numerical confirmation
%(with Level Set Method) of Mullins' analytical
%results for an isolated groove with small slopes of the groove profile
%at the
%groove root \cite{MULLINS57}; {\bf b)}
%to obtain some predictions for the evolution of
%an isolated groove with large slopes at the groove root.
%{\bf c)} to demonstrate the interaction of two adjacent grooves in a
%periodic array.

Below we give more details related to the physical model.

\begin{figure}[H]
\centering
\psfig{figure=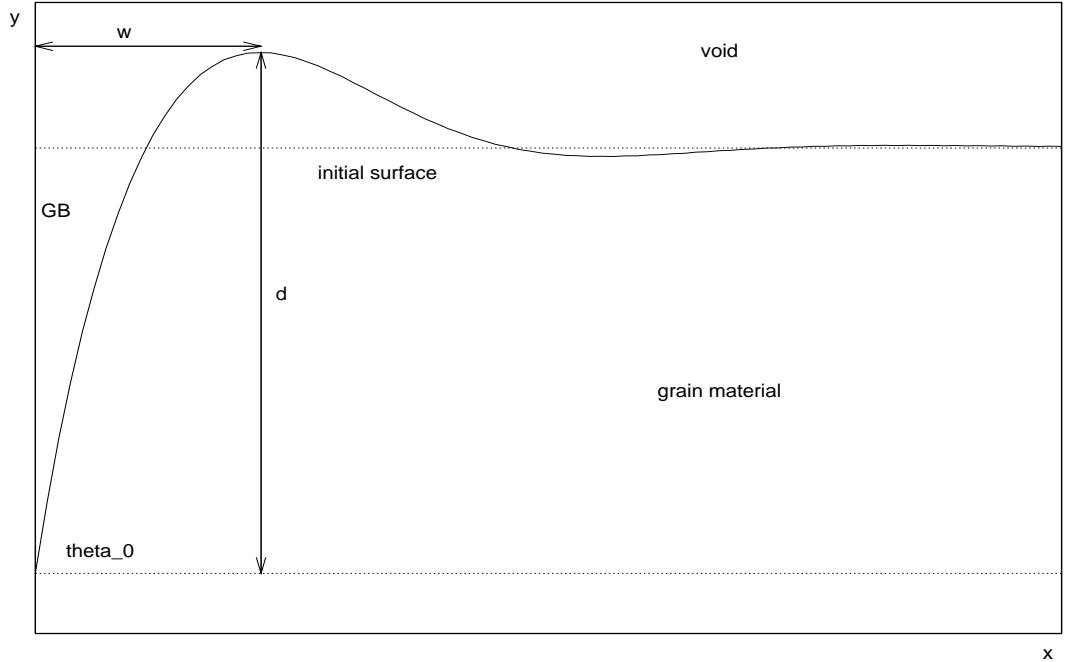,height=3.5in,width=5.5in,angle=270}
\caption{Sketch of a GB groove. $w$ denotes the half-width of the
groove, and
$d$ denotes
the depth.}
\label{Fig1}
\end{figure}

\begin{itemize}
\item {\bf Driving Forces and Diffusion Mobilities}

In the absence of an electric current, the diffusion is
driven by a variation in chemical potential, $\mu$, which causes atoms
to migrate
from high potential to low potential regions.
It may be shown that \cite{MULLINS57}
\begin{equation}
\label{1.1}
\mu(K)=K\gamma\Omega,
\end{equation}
where $K$ is the surface curvature,
$\gamma$ is the surface tension, and $\Omega$ is the
atomic volume. Gradients of chemical potential are therefore associated
with gradients of curvature.

In interconnects, GBs represent numerous fast diffusion
pathways
with high diffusion coefficient, $D$. As a matter of fact, the bulk
diffusion
can be neglected \cite{MULLINS57}.
The diffusion flux along the GB, $I_{GB}$, is given by
\begin{equation}
\label{1.2}
I_{GB} = \frac{D \delta}{k T}  \nabla \mu,
\end{equation}
where $\delta \sim 10^{-8} cm$ is the GB thickness, $k$ the Boltzmann
constant and
$T$ the absolute temperature.

Let ${\bf{\tau}}$ be the tangential direction to the surface profile in
2D.
If ${\bf n}=(n_x, n_y)$ is the unit vector normal to the surface or GB,
then the
following relations hold:
\begin{equation}
\label{1.3}
{\bf{\tau}} = (n_y, -n_x), \quad
\frac{\partial K}{\partial \tau}  =  {\bf \nabla K} \cdot {\bf{\tau}} =
\frac{\partial K}{\partial x} n_y - \frac{\partial K}{\partial y} n_x
\equiv K_{\tau}.
\end{equation}
The surface diffusion flux along the groove walls is given by the
formula
\begin{equation}
\label{1.4}
J  =  -B \ K_{\tau},
\end{equation}
where
\begin{equation}
\label{1.5}
B  =  \frac{D \delta \gamma \Omega^{4/3}}{k T}
\end{equation}
is known as Mullins' constant.
Note that $J$ is proportional to the first directional
derivative of the curvature.

\item {\bf Boundary Conditions}

The boundary conditions at the groove root are dictated by the local
equilibrium between the surface tension, $\gamma$, and the GB tension,
$\gamma_{gb}$. In the symmetric case of the GB ($x=0$) normal to an
original ($y=const.$) flat surface, the angle of inclination of the
right branch
of the surface at the groove root with respect to the $x$ axis is
$\theta_0 = sin^{-1}(\gamma_{gb}/2\gamma)$  (see Fig. \ref{Fig1}).

The rapid establishment of the equilibrium angle between the GB
and the surface by atomic migration in the vicinity of the intersection
develops some curvature gradient at the adjacent surface
and thus
induces a surface diffusion flux along the groove wall
in the direction away from the groove
root, opposite to the groove extension direction.

Other boundary conditions depend on the particular problem (presence or
absence
of GB flux, etc.)

\end{itemize}

\Section{Mathematical Model}
\subsection{The Conventional Approaches}

An adequate mathematical model which captures the above physical
phenomena
in interconnects was developed first by Mullins \cite{MULLINS57} and
extended by him and others \cite{KGFMB1,KGFMB2,MILLINS}.
It
describes the evolution of the groove shape, $y(x,t)$, and has the form
of a transport equation
\begin{equation}
\label{2.1}
y_t  =  -J_x = -B \left\{ (1+y_x^2)^{-1/2}
\left[ (1+y_x^2)^{-3/2}y_{xx} \right]_{x} \right\}_x.
\end{equation}
$J$ and $B$ are given in \rf{1.4} and \rf{1.5}.

For an isolated GB at $x=0$, the groove continues to develop
because the material
continues to move from the curved shoulder of the groove to the flat
surface.
The classical description is provided by an analytic solution (on the
$x>0$ side)
of the linearized version
of the equation \rf{2.1} (the ``small slope approximation", SSA).
The linearized equation has the form \cite{MULLINS57}
\begin{equation}
\label{2.2}
y_t  =  -B\ y_{xxxx},
\end{equation}
subject to the initial condition
\begin{equation}
\label{2.3}
y(x,0)=const,
\end{equation}
and the boundary conditions
\begin{equation}
\label{2.4}
y_x(0,t)=\tan{\theta_0}=m \ll 1,
\end{equation}
$$
J(0,t)=y_{xxx}(0,t)=0,
$$
$$
y(x \rightarrow \infty, t) = const \quad \mbox{with all derivatives}.
$$
The first condition in \rf{2.4} is the small slope approximation itself.
The second one reflects the absence of a GB flux $I_{GB}$.
The solution describes a profile with a constant shape whose size
is increasing  all the time.

Although this analytical approach describes some basic phenomena in
interconnects,
it is of limited use due to the restriction on the steepness of the
slope.
There are several numerical techniques which are widely used in modeling
moving fronts,
such as the {\em
marker/string} (M/S)
methods \cite{SETHI} or the {\em volume-of-fluid} (VOF) methods
\cite{CHORIN,NW}.
These methods deal directly with the evolution equation of type
\rf{2.1}, and
therefore are ``explicit" methods. The M/S
methods come from Lagrangian approach to front evolution
problems. In the Lagrangian approach, the grid is attached to the moving
front. A known drawback of the Lagrangian approach is that it is not
well-suited for the computation of bifurcating fronts. Besides,
stability and local singularity problems are more emphasized in these
methods than in methods based  on Eulerian approach, such as the VOF
method.
The Eulerian approach,
in which the front moves through a grid which is fixed in space, does
not have these drawbacks, but - as it is known - here the fronts are
diffused.
In addition, some intricate (subcell) bookkeeping is required
to properly keep track of fronts.

There are numerical approaches which are based on {\em finite-element}
discretization
of the computational region \cite{BOWER-FREUD}. However, they result in
complicated algorithms which involve many computational steps such as
computations of the following: displacement field of material points
from a reference configuration, the stress field as a result of
diffusion in the
solid and geometry update of interfaces.
Besides, the computational complexity grows since higher resolution is
required as the shape of the interface becomes more complicated.
As a result, these methods are unable to handle very complex
multidimensional
boundary shapes.

\subsection{The Proposed Solution: Usage of the Level Set Method}

To ``capture" the interface (rather than to track it),
our method of choice is the ``implicit" LS method.
The method was introduced by Osher and Sethian and was further
developed during the last several years (for an introduction to the LS
methods and
an exhaustive bibliography list see the  monographs by
Sethian \cite{SETYAN-BOOK1,SETYAN-BOOK2}).
The method enables to capture drastic changes in the
shape of the curves (interfaces) and even topology changes.
% As pointed out in
%\cite{PMOZK},
%``the advantages of this capturing approach are well known by now. The
%method is
%stable, the equations are not unnecessarily stiff, geometric quantities
%such
%as curvature become easy to compute, and 3D problems present no
%difficulties".

The basic idea of the method consists of embedding the curve $y(x,t)$
into a higher dimensional space. As a matter of fact, we consider the
evolution of a two-dimensional field $\phi(x,y,t)$ such that its zero
level set, $\phi(x,y,t)=0$, coincides with the curve of interest,
$y(x,t)$, at any time moment $t$. The level set function $\phi(x,y,t)$
can be
interpreted as a signed distance from the curve $y(x,t)$, which moves in
the
direction normal to itself.

The evolution of $\phi(x,y,t)$ is described by an Hamilton-Jacobi
type equation. A remarkable trait of the method is that the function
$\phi(x,y,t)$ remains smooth, while the level surface $\phi = 0$ may
change
topology, break, merge, and form sharp corners as $\phi$ evolves. Thus,
it is possible to perform numerical simulation on a discrete grid
in the spatial domain, and substitute  finite difference approximation
for the spatial and temporal derivatives in time and space.
%Another nice feature of the method is that the explicit location of the
%interface
%need not to be known in the computational process; all the necessary
%information
%is extracted from the level set function.

The evolution equation has the form
\begin{equation}
\label{2.5}
\phi_t + F |\nabla \phi| =  0, \quad \mbox{given}\ \ \phi(x,t=0).
\end{equation}
%
%where $|\nabla \phi| =  \sqrt{ \phi_x + \phi_y }$.
The normal velocity, $F$, is considered to be a function of spatial
derivatives of $\phi(x,y,t)$. In many applications $F$ is a function of
the curvature, $K$, and its spatial derivatives. The curvature $K$ may
be computed
via the level set function $\phi$ as follows:
\begin{equation}
\label{2.6}
K  =  \nabla \cdot {\bf n}, \quad {\bf n}  =
\frac{ \nabla \phi}{|\nabla
\phi|}=\left(\frac{\phi_x}{\left(\phi_x^2+\phi_y^2\right)^{1/2}},
\frac{ \phi_y}{\left(\phi_x^2+\phi_y^2\right)^{1/2}}\right).
\end{equation}
Here ${\bf n}$ is \lq \lq normal vector", and it coincides with the
(previously introduced)
unit
normal to the surface, $y(x,t)$, on the zero level set $\phi=0$.
Formulas \rf{2.6} can be combined as follows
\begin{equation}
\label{2.7}
K  =  \nabla \cdot \frac{ \nabla \phi}{|\nabla \phi|} =
\frac{\phi_{xx} \phi_y^2 - 2 \phi_{x} \phi_y \phi_{xy} + \phi_{yy}
\phi_x^2}
{ \left( \phi_{x}^2 + \phi_y^2 \right)^{3/2}},
\end{equation}
and the sign of $K$ is chosen such that a sphere has a positive mean
curvature
equal to its radius. In the case of surface diffusion in 2D,
\begin{equation}
\label{2.8}
F  =  -B K_{\tau \tau}.
\end{equation}

One drawback of the LS method stems from its computational expense.
Its complexity seems to be as much as $O(n^2)$ operations per time step,
which is more than any Lagrangian method which necessitates
$O(n)$
operations per time step, where $n$ is the number of grid points in the
spatial
direction.
It is possible, however, to reduce the complexity of the LS method to
$O(n)$
using a local (another term is narrow band (tube))
approach \cite{SETYAN-ADALS1,PMOZK}.
This is achieved by the construction of an
adaptive mesh around the propagating interface. We distinguish
between
the ``near field", which is a thin
band of neighboring level sets around the propagating front,
 and the ``far field" which contains
the rest of the  grid points.
The evolution equation is solved only in the near field.
The values of $\phi$ at grid points in the far field are not updated
at all. When the interface in motion reaches the edge of the narrow
band, a new narrow
band is built around the current interface position. Note that this
could be done
without interface reconstruction from the level set function (which
requires some additional computations).
We just have
to examine the shift in the sign of $\phi$ at grid points adjacent
to the interface.
The width of the narrow band is determined as a balance between the
computation
involved in the  re-built and the
calculations performed on far away points.

In most of the applications of the LS method to date, the driving forces
were
proportional to the curvature (see \cite{SETYAN-BOOK1,SETYAN-BOOK2} for
review
and discussion). There are only few applications
\cite{AIR,CHOPP-SETYAN,LZG} where the
driving force is proportional to the {\em second directional derivative}
of the curvature
(in the 3D case, to the surface Laplacian of curvature which is
constructed
from the derivatives in each principal direction),
which is the case for the normal velocity function \rf{2.8}. Therefore,
the present materials science
problem presents a rather new (from the mathematical
point of view) application for the LS method.
As pointed out in \cite{CHOPP-SETYAN}, ``this is an intrinsically
difficult problem
for three reasons. First, owing to the lack of a nice maximum principle,
an embedded curve need
not stay embedded, and this has significant implications in attempting
to analyze motion which
results in topological change. Second, the equations of motion contain a
fourth derivative term,
and hence are highly sensitive to errors. Third, this fourth derivative
term leads to
schemes with very small time steps."

\subsection{Computational Algorithm}

A typical computational domain
is a rectangular box $[0,l_1; 0,l_2]$ of a material in 2D.

\par\noindent
The proposed computational algorithm consists of the following steps:
\vspace{0.5cm}\\
%\bigskip
{\bf BEGIN ALGORITHM}

\begin{description}

\item[1. Discretization - ]  The entire computational region $W$ is
discretized using a
uniform grid $x_i = i \Delta x, y_j = j \Delta y, i=0...N, j=0...M$,
where $N$ and $M$ are the number of grid points in $x$- and $y$- directions
respectively.
The functions are projected on this grid, so that
 $\phi(x,y,t)=\phi_{i,j}(t)$.

\item[2. Initialization - ]
The initial interface, $y(x,t=0)$, is defined analytically, or as a set
of points in $W$
(the points lie on $x=const$ grid lines, but not necessarily on
$y=const$ grid lines).
In the latter case, we define a cubic spline $\xi(x,t=0)$ passing
through these points in order
to be able to perform further initializations. The function $y(x,t=0)$
needs
not to be necessarily smooth (i.e., it may feature sharp corners,
discontinuities, etc.),
but, in our implementation it must be single-valued in order to make it
possible to choose the sign of $\phi$ (below).
This is because we are only interested in the particular case of
analyzing the motion
of open curves which may be described by functions during the whole
process of the
evolution.

We also define the near field and the far field.
The width of the near field is usually 5 to 10 grid levels (points).

In the region $W$, the level set function $\phi$ is initialized as
an exact signed
distance function to the initial interface (see Fig. \ref{Fig2}),
\beq{2.9}
\phi(x_i,y_j,t=0) < 0 \quad \mbox{if} \quad y_j < y(x,t=0)
\eeq
$$
\phi(x_i,y_j,t=0) = 0 \quad \mbox{if} \quad y_j = y(x,t=0)
$$
$$
\phi(x_i,y_j,t=0) > 0 \quad \mbox{if} \quad y_j > y(x,t=0).
$$

\noindent
%Note that our choice of the sign of $\phi$ is consistent with
%the direction of the normal vector ${\bf n}$ \cite{CHOPP} (see Fig.
\ref{Fig2}).
Since $\phi(x,y,t=0)$ is a signed
distance function, then $|\nabla \phi(x,y,t=0)|=1$.

\item[3. Compute ] normal vector components and curvature using formulas
\rf{2.6}, \rf{2.7}.
The derivatives in \rf{2.6}, \rf{2.7}
(as well as in other functions of $x,y$ except the gradient term
in the evolution equation itself, see {\bf step 6}) are discretized
using the standard second order
accurate central difference approximations. Fourth-order accurate
approximations were tested also
but we did not observe any particular
increase in the global accuracy of the calculations. In addition, in
this
case, the implementation
of the boundary conditions with the level set function is
problematic due to the use of a  wide
stencil. The time step also needs to be reduced in order to have
stability. We
find that
the standard central difference scheme works well for us.

\begin{figure}[t]
\centering
\psfig{figure=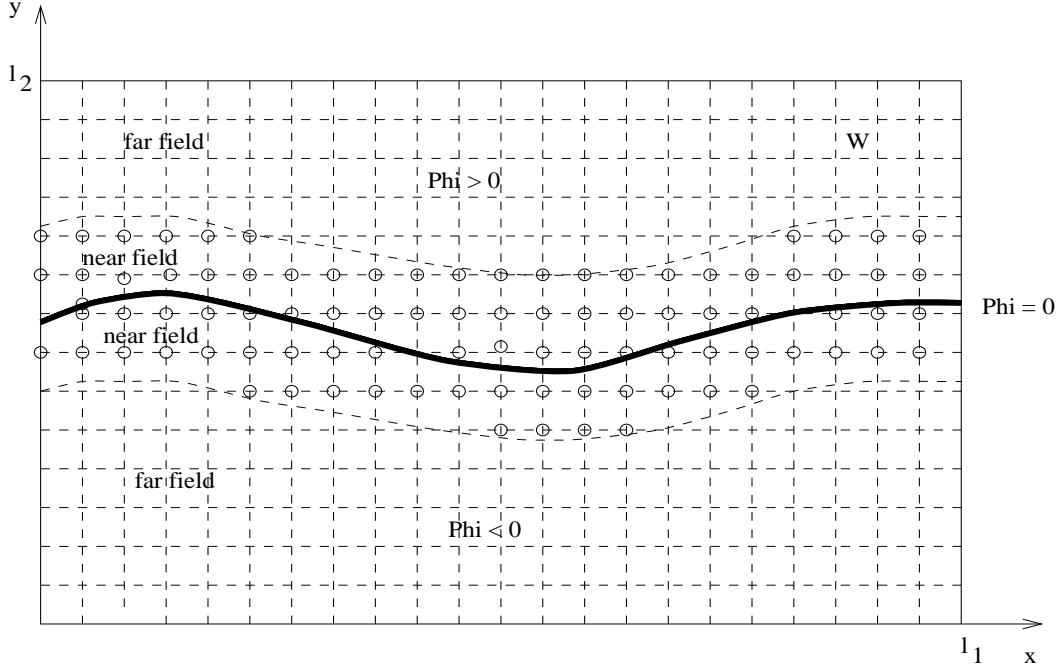,height=3.5in,width=5.5in,angle=270}
\caption{Computational domain.}
\label{Fig2}
\end{figure}

\item[4. Compute ] first directional derivative of the curvature,
$K_\tau$,
using
the formula \rf{1.3} and second directional derivative of the curvature,
$K_{\tau \tau}$,
\beq{2.10}
K_{\tau \tau} = \nabla\left[ \nabla K \cdot {\bf \tau} \right]\cdot {\bf
\tau} =
\frac{-K_{xx} \phi_y^2 + 2 K_{xy} \phi_x \phi_{y} - K_{yy} \phi_x^2}
{ \phi_{x}^2 + \phi_y^2} +
\frac{K\left(K_{x} \phi_x + K_{y} \phi_y\right)}
{ \left( \phi_{x}^2 + \phi_y^2 \right)^{1/2}}=
\eeq
$$
\frac{-K_{xx} \phi_y^2 + 2 K_{xy} \phi_x \phi_{y} - K_{yy} \phi_x^2}
{ \phi_{x}^2 + \phi_y^2}+
K\left[K_\tau+K_y(n_x+n_y)-K_x(n_y-n_x)\right].
$$
We now have  the normal velocity function \rf{2.8} and the flux
\rf{1.4}.

\item[5. Choose ] time step. The CFL condition for the surface diffusion
is
\beq{2.11}
\Delta t_1 \leq \mbox{min}^{4}(\Delta x, \Delta y)/B.
\eeq
The CFL condition for the Hamilton-Jacobi equation in updating the
velocity is
\beq{2.12}
\Delta t_2 \leq \mbox{min}(\Delta x, \Delta y)/F_{max},
\eeq
where $F_{max}$ is the largest magnitude of the normal velocity in the
computational domain.
%The CFL condition for the surface diffusion due to applied
%*electric field is, obviously (??)
%\beq{2.11}
%\Delta t_3 \leq \mbox{min}^{2}(\Delta x, \Delta y)/C.
%\eeq
%
The adaptive time step $\Delta t$ is chosen as the smallest of the two.
%Usually $B$ is relatively small compared with $C$, and the time step is
%tolerable
%for the simulations on most workstations and even PCs.

\item[6. Compute ] backward and forward gradient functions; {\bf update}
$\phi$ from the
evolution equation using explicit time-stepping scheme. The solutions of
equation
\rf{2.5} are often only uniformly continuous with discontinuous
derivatives,
no matter how smooth the initial data is \cite{OSHER-SETYAN,OSHER-SHU}.
Simple central differencing is not appropriate here to approximate the
spatial derivatives
in $|\nabla \phi|$.
Instead, we use Essentially Non-Oscillatory (ENO) type schemes for
Hamilton-Jacobi equations
as developed in
\cite{OSHER-SETYAN,OSHER-SHU,SHU-OSHER}. More precisely,
we
use second-order ENO scheme given explicitly in \cite{ZCMO}. To update
$\phi$ for one time step,
the simplest method is to use Euler, i.e.
\beq{2.13}
\phi^{n+1}=\phi^n+\Delta t L(\phi^n),
\eeq
where $L(\phi)$ is the spatial operator in \rf{2.5}.

\item[7. Update ] near field. Check the sign of $\phi$ at the grid
points adjacent to
the interface and compute the new locations of near field points.

\end{description}

{\bf Go to step 3}

\medskip
{\bf END ALGORITHM}
\bigskip

\par\noindent
{\bf Remark 1:} To achieve a uniformly
high-order accuracy in time, we replace \rf{2.13} with the second-order
Total Variation
Diminishing (TVD)
Runge-Kutta type discretization \cite{OSHER-SHU,SHU-OSHER},
which reads
\beq{2.14}
\tilde \phi^{n+1}=\phi^n+\Delta t L(\phi^n)
\eeq
$$
\phi^{n+1}=\phi^n+{\Delta t \over 2} \left[L(\phi^n)+L(\tilde
\phi^{n+1})\right]
$$
The necessary changes to the algorithm are obvious. The choice
of such a low-order
Runge-Kutta scheme is justified by the fact that the time step, dictated
by stability
requirements, is very small.

\par\noindent
{\bf Remark 2:} It is highly desirable that the level sets behave
nicely, in the sense that
two different level sets do not cross, and in fact remain roughly evenly
spaced in time.
In terms of the level set function $\phi$, this corresponds to the fact
that
the gradient of $\phi$ at any given
point of a level set does not change dramatically over time.
For the numerical method this translates into numerical stability.
The best way to achieve this is to keep $\phi$ close to the signed
distance function
(or even to keep it exactly equal to the signed distance function), thus
keeping
$|\nabla \phi| \approx(=) 1$.
The operations performed on $\phi$ that accomplish it are called
``reinitialization".
To summarize, reinitialization is the process of replacing $\phi(x,y,t)$
by another
function $\tilde \phi(x,y,t)$ that has the same zero contour as
$\phi(x,y,t)$ but behaves
better,
and then taking this new function $\tilde \phi(x,y,t)$ as the initial
data to use until the next
round of reinitialization.
There are several ways to do this. The straightforward one (first
proposed in \cite{MSV} and
recently used in \cite{AIR}) is to interrupt the time-stepping,
reconstruct the
interface using some interpolation technique and directly compute a new
signed distance function to the interface.
This approach is very expensive and also may bring some undesirable
side effects, such as oscillations in the curvature. Instead, we use the
iteration procedure
of \cite{SSO}.
The function $\phi$ is reinitialized
by solving the following Hamilton-Jacobi type equation to its steady
state, which is the desired
signed distance function:
\beq{2.15}
\phi_t
 = S(\phi_0) \left( 1 - |\nabla \phi| \right),
\eeq
where $S$ is a smoothed sign function
\beq{2.16}
S(\phi_0) =  \frac{ \phi_0}{ \sqrt{ \phi_0^2 + \epsilon^2} }, \quad
\epsilon = \mbox{min}(\Delta x, \Delta y).
\eeq
The same second-order ENO and TVD Runge-Kutta schemes used for the
solution of the
equation \rf{2.5} are used for the iteration of \rf{2.15}.
As a rule, three or four iterations are sufficient to evolve $\phi$
close enough
to the desired signed distance function.
An important practical question is how frequently the reinitializations
are applied.
In some applications of the level set method the reinitializations could
be triggered
after a fixed number of time steps. However, we achieved the best
results by reinitializing
every time step in the band of level sets that contains points from the
near field.

\par
\begin{figure}[t]
\centering
\psfig{figure=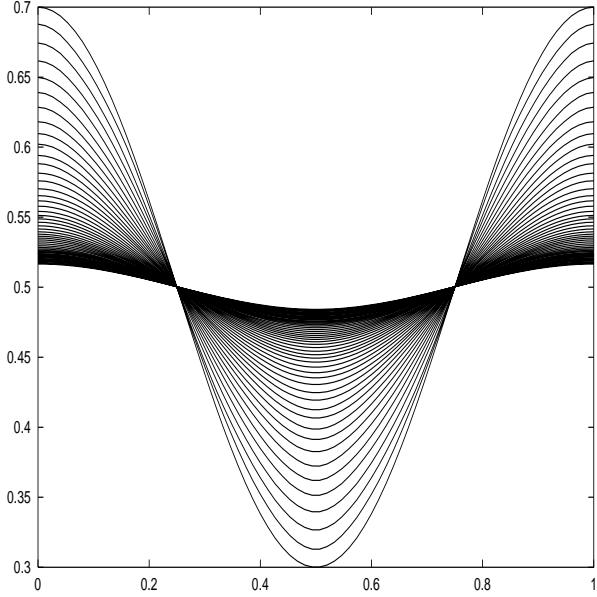,height=3.25in,width=3.25in,angle=270}
\caption{The cosine curve, evolving
under \rf{2.5} with $F=-0.1\ K_{\tau\tau}$. A coarse 75$\times$75
grid is used.
25000 time steps were made by the Runge-Kutta  integrator (the shape is
printed out every 500 steps),
and we reinitialize in every step.}
\label{Fig3}
\end{figure}

\begin{figure}[H]
\centering
\psfig{figure=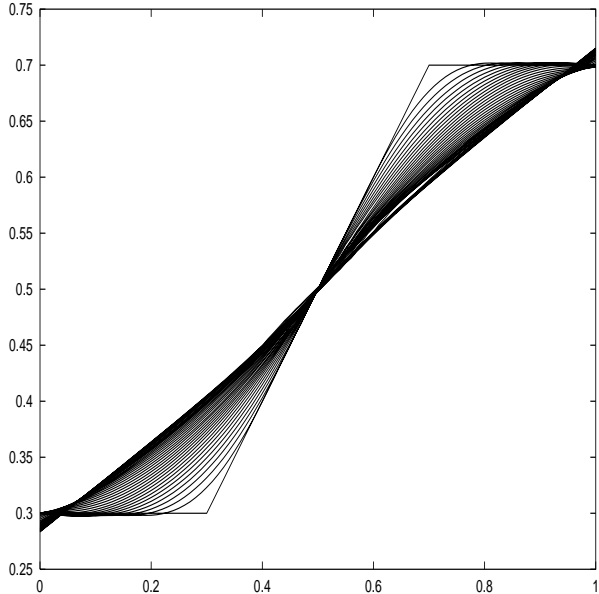,height=3.25in,width=3.25in,angle=270}
\caption{The evolution of a non-smooth curve (step function). The grid
used is
100$\times$100,
20000 time steps were made.}
\label{Fig4}
\end{figure}

\noindent
{\bf Remark 3:}  The evolving interface touches the vertical
boundaries $x=0,x=l_1$
by its ends and therefore any boundary conditions imposed on vertical
walls
influence the evolution of the front.
This is why, depending on the nature of the problem, we either choose
periodic b.c.
at vertical walls or just an approximation of  the derivatives
at vertical walls by one-sided differences. At the horizontal walls, we
always
use one-sided differences.
For illustration purposes, in Fig. \ref{Fig3} we present part of the
cosine
curve evolving
under \rf{2.5} with the speed function $F=-0.1\ K_{\tau\tau}$.
Boundary conditions  at vertical walls are periodic. Note that the speed
of evolution
slows as the curve approaches equilibrium state with $K=0$ (line
$y=0.5$). This is because the
curvature, and hence its derivative,  become smaller.
In order to demonstrate the abilities of the method, in Fig. \ref{Fig4}
we
present the
evolution of a non-smooth curve (step function) under the same speed
law.

\par\noindent
{\bf Remark 4:} The very special feature of the presented implementation
of the
Level Set Method is the incorporation  of physical boundary conditions
into
the Level Set
numerical scheme. Most of the implementations known so far lack this
complication. Usually only closed interfaces far away from any
boundaries domains  are
considered while the evolution proceeds far away from the boundaries.

For the GB grooving by surface diffusion, two boundary conditions at the
groove root are essential: these are conditions of type \rf{2.4},
reflecting
the fixed slope of the interface and the absence of GB atomic flux.
The boundary conditions we impose at $x=l_1$ are zero slope of the
interface and
zero flux. The first condition echoes the initial flat interface.
The second condition guarantees the conservation of matter, i.e. a
constant
area under the groove profile during the evolution.

%It is noteworthy that the presented algorithm conserves the area
%perfectly.
%If we define $S_u$ to be the ares above the curve, $S_b$ to be the area
%below the curve,
%then, if the method is conservative, the ratio $S_u/S_b$ must remain
%equal to
%$(S_u/S_b)_{t=0}$ for all times. This property holds nicely for the
%motion depicted in
%Figs. \ref{Fig3} and \ref{Fig4} and
%the area losses  do not exceed 1\%.

%The choice of boundary conditions
%at horizontal boundaries $y=0,y=l_2$ is not so crucial, since boundary
%points constitute
%the far field.  At $y=0,y=l_2$ we require the continuity of $\phi$ in
%the normal direction
%or, alternatively, we approximate the derivatives in the $y$-direction
%by unilateral
%differences, as done in \cite{AIR}. For example, the second-order
%accurate unilateral finite
%difference approximation for the central first and second derivatives
%read
%\beq{2.16}
%y=0: \quad {\partial \phi \over{\partial
%y}}={-3\phi_{i,0}+4\phi_{i,1}-\phi_{i,2} \over {2\Delta y}},
%\quad {\partial^2 \phi \over{\partial
%y^2}}={2\phi_{i,0}-5\phi_{i,1}+4\phi_{i,2}-\phi_{i,3}\over {(\Delta
%y)^2}},\ i=0..N,
%\eeq
%\beq{2.17}
%y=l_2: \quad {\partial \phi \over{\partial
%y}}={3\phi_{i,M}-4\phi_{i,M-1}+\phi_{i,M-2} \over {2\Delta y}},\
%{\partial^2 \phi \over{\partial
%y^2}}={2\phi_{i,M}-5\phi_{i,M-1}+4\phi_{i,M-2}-\phi_{i,M-3} \over
%{(\Delta y)^2}},
%\eeq
%
%where $N+1$ and $M+1$ are the numbers of grid points in horizontal and
%vertical directions respectively.

Special attention was given to the treatment of these boundary
conditions
within the framework of the Level Set method.
Two methods were developed.
\begin{description}
\item[The simplest technique ] is the use of correction step in the
iterative algorithm.
The fixed slope at the groove root is achieved
in the following way:
at every time step, the interface is reconstructed from the $\phi$-field
and the locations of the two end-points of the interface (at $x=0$ and
$x=l_1$,
respectively) are corrected in order to preserve the small-slope
and the
zero-slope conditions. Then, for all grid points that lie on grid lines
$x=0$ and $x=l_1$, it is sufficient to directly compute a  new signed
distances to the updated locations of the interface end points. This way
we
incorporate
the new locations of the end points back into the $\phi$-field.
This direct reinitialization is performed only for a few grid
points
that lie on vertical boundaries and, besides, this computation does not
contain iteration loop.
The zero flux conditions could be imposed locally, i.e. in the vicinity
of the
groove root and of the interface end-point at $x=l_1$,
or along the the entire $x=0$ and $x=l_1$ grid lines. After
the computed values
of $K_\tau$ are reset to zero, the
$K_{\tau\tau}$
is computed according to eq. \rf{2.10}, where $K_\tau = 0$ at
$x=0,l_1$ and
$K_\tau \neq 0$ otherwise. After multiplication by $-B$, this gives the
values of the
normal velocity function \rf{2.8}, corrected by the zero flux
constraint.
\item[Extension of the $\phi$-field beyond the GB] makes use of Taylor
expansion
up to second order, as follows (also see eq. \rf{2.6}):
\beq{2.17}
\phi_{-1,j}
 = \phi_{0,j} - \phi_x\left|_{0,j}\right.\Delta x =
\phi_{0,j} - |\nabla \phi_{0,j}| \ n_x\left|_{0,j}\right.\Delta x =
\phi_{0,j} + |\nabla \phi_{0,j}|\sin{\theta_0}\ \Delta x,
\eeq
where $\phi_{-1,j}$ is one grid point beyond the GB.
Equation \rf{2.17} incorporates the groove root angle.
Then we compute in \rf{2.7} the curvature values, $K_{0,j}$, along the
GB,
using both the values of $\phi$ inside the computational domain
($\phi_{1,j}$) and outside ($\phi_{-1,j}$). This also gives us the
values of $K_y\left|_{0,j}\right.$. The zero flux condition is applied
using equation
\rf{1.3} which, after substitution of normal vector components from
\rf{2.6}
and rearrangement of the terms become
\beq{2.18}
K_x\left|_{0,j}\right.
 = \frac{K_\tau |\nabla \phi|+K_y
\phi_x}{\phi_y}\left|_{0,j}\right.=-K_y\left|_{0,j}\right.
\tan{\theta_0}.
\eeq
Applying Taylor expansion again, we get the ghost values of the
curvature:
\beq{2.19}
K_{-1,j}
 = K_{0,j} - K_x\left|_{0,j}\right.\Delta x,
\eeq
where $K_x\left|_{0,j}\right.$ is given by \rf{2.18}. Now all the data
is known
and we can compute
the values of $K_{\tau \tau}$ from \rf{2.10} and the values of the
normal velocity from
\rf{2.8}.
\end{description}
Both methods were successfully used in calculations.

\Section{Numerical results and discussion}

\begin{figure}[H]
\centering
\psfig{figure=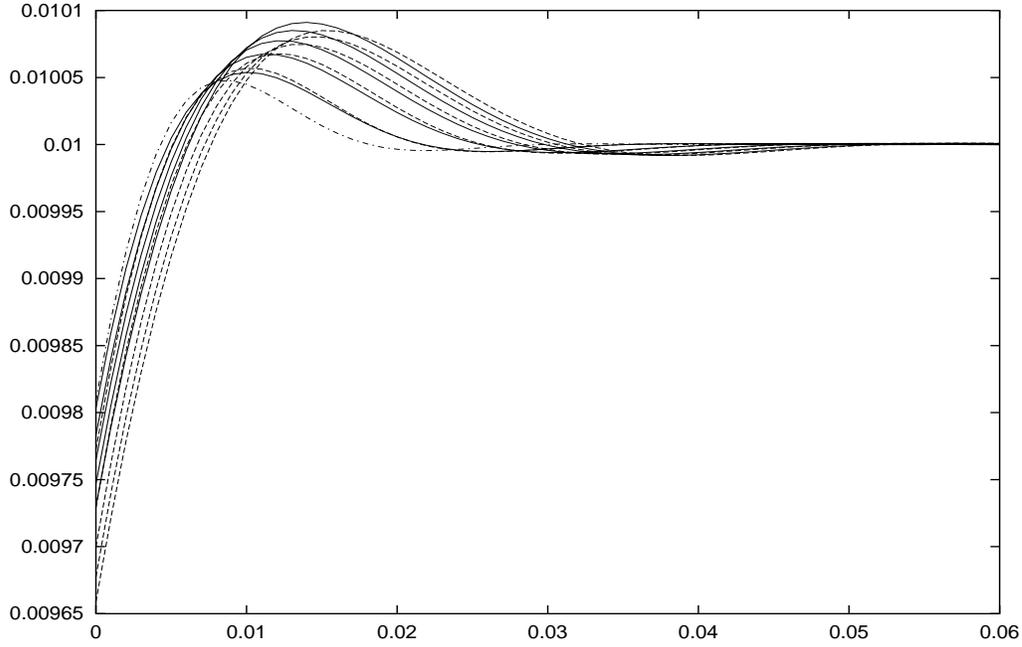,height=3.5in,width=5.5in,angle=270}
\caption{GB grooving by surface diffusion. The slope at groove root
is $m=6.55e-02$. The initial interface is shown with dashed-dotted line,
the numerical results obtained by means of the LS Method
are shown with solid lines, the reference
results of \cite{MULLINS57} are shown with dashed lines.}
\label{Fig5}
\end{figure}

\begin{figure}[H]
\centering
\psfig{figure=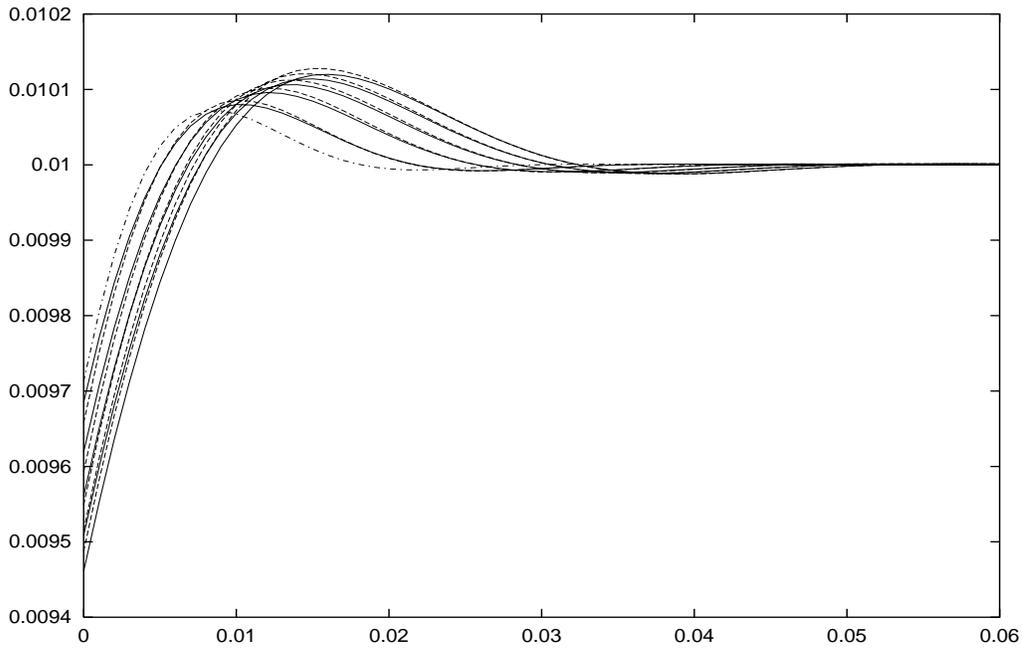,height=3.5in,width=5.5in,angle=270}
\caption{Same as Figure \ref{Fig5}, but the slope at groove root
is $m=9.85e-02$.}
\label{Fig6}
\end{figure}

\begin{figure}[H]
\centering
\psfig{figure=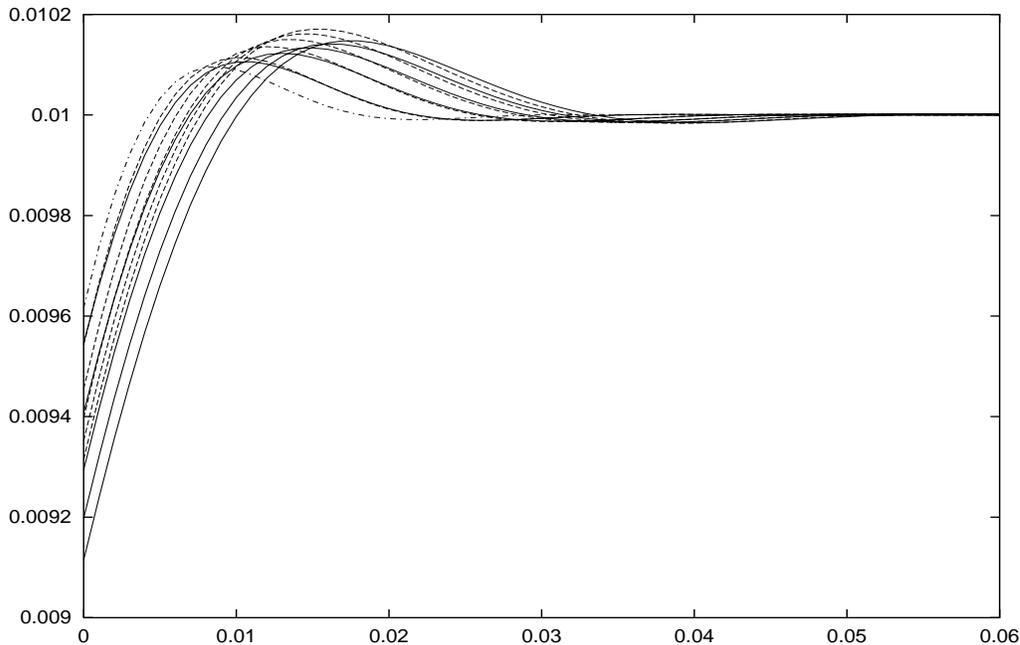,height=3.5in,width=5.5in,angle=270}
\caption{Same as Figures \ref{Fig5}, \ref{Fig6}, but the slope at groove
root
is $m=1.32e-01$.}
\label{Fig7}
\end{figure}

Figures \ref{Fig5} to \ref{Fig7} show the groove profile having
different slopes at the groove
root, evolving
under \rf{2.5} with a speed function $F= -BK_{\tau\tau}$.
We take $B=0.025$.
The profile is symmetric with respect to the GB at $x=0$, therefore only
its
right
part is calculated.

\noindent The results obtained by means of the LS Method are shown with
solid lines, while reference
results
for Mullins' problem \rf{2.2}-\rf{2.4} are shown with dashed lines.
In all the three numerical experiments reported here the dimensions of
the computational
box are $[0.,0.08; 0.,0.02]$, the mesh is 120$\times$40.

Our initial interface for the Level Set simulations already
has the shape of Mullins' groove.
The reason we don't have a flat interface $y(x,0)=const.$ as an initial
condition
is that
the LS formulation requires a non-zero initial curvature, otherwise the
curve does not
evolve at all (since $F=0$ in this case). The initial interface in
Figs. \ref{Fig5} - \ref{Fig7} is shown with dashed-dotted line.

The initial Mullins' groove
is obtained as follows:
we integrate numerically the equation \rf{2.2} using the method-of-lines
approach.
The time integrator is second-order Runge-Kutta and the spatial operator
is
discretized using
second order central differences. The integration proceeds from $t=0$
to
$t=8.0e-09$. The initial and boundary conditions are \rf{2.3} and
\rf{2.4}, where
$\theta_0=\pi/48,\pi/32,\pi/24$ stands for Figs. \ref{Fig5} -
\ref{Fig7},
respectively. The corresponding slopes are
$m=6.55e-02,9.85e-02,1.32e-01$.
The practical values used in experiments lie
between 0.05 to 0.2 and
the range of the groove depth in experiments is between 0.1$\mu$ and
1$\mu$.
The reason we anticipate the use of the analytic solution to the
Mullins' problem \rf{2.2}-\rf{2.4} (either it exists) is the truncation
of
infinite series
in which this solution is represented. The reference results for later
times
are also obtained using the described numerical procedure.

In \cite{MULLINS57}, two kinetic laws were established (within the
framework
of the SSA). One concerns the evolution
of the depth of the groove with respect to the maximum surface elevation
(see Fig. 1).
The depth, $d$, is governed by
\beq{3.1}
d=0.973\ m\ (Bt)^{1/4}.
\eeq
The other kinetic law concerns the evolution of the distance between the
position
of the groove root and that of the surface maximum. In the case of the
symmetric groove,
we call it the half-width, $w$, of the groove. It is governed by
\beq{3.2}
w=2.3(Bt)^{1/4}.
\eeq
From these expressions, we have the time independent ratio
\beq{3.3}
w/d=2.3515/m.
\eeq

Under typical experimental conditions a groove of depth $d=0.3\mu$
is formed within $t=10^4$ sec (2.4 hr).
It is shown in \cite{MULLINS57}, that
it would require approximately 8 days to triple this depth.
This explains why in our numerical experiments the groove seems to stop
developing at later times. The physical reason for this is the increase
in
the length of a path along which the surface diffusion takes place.
As a rule, we stop the run when the groove doubles its depth or width.

For the slopes considered, we observe good qualitative agreement
with Mullins' solution. The small difference is due to two reasons.
First, the results to which we compare are obtained by integrating
the linearized equation \rf{2.2}, which is, strictly speaking, valid
only for infinitesimal slopes. The slopes we choose are, of course,
finite,
and the governing equation we solve, i.e., the equation \rf{2.5} is
fully nonlinear.
Second, there are
inevitable area losses, since the LS method is not fully conservative.
For bigger slopes, our grooves appear to be deeper and wider than
Mullins' one.

In Tables 1 to 3, the results for all three
tests are summarized.

\vspace{0.5cm}
\begin{center}
Table 1. Our results for GB grooving, compared with classical Mullins'
results.
The slope at groove root is $m=6.55e-02$.
\vspace{0.5cm}

\begin{tabular}{|c||c||c||c||c||c||c||c|}
\hline
\multicolumn{1}{|c|}{ {step}  } &
\multicolumn{1}{|c|}{ {$t$}  } &
\multicolumn{1}{|c|}{ {$d$,eq.\rf{3.1}} } &
\multicolumn{1}{|c|}{ {$d$,LS M.} } &
\multicolumn{1}{|c|}{ {$w$,eq.\rf{3.2}} } &
\multicolumn{1}{|c|}{ {$w$,LS M.} } &
\multicolumn{1}{|c|}{ {$w/d$,eq.\rf{3.3}} } &
\multicolumn{1}{|c|}{ {$w/d$,LS M.} }\\ \hline \hline
0 & 8.0e-9 & 2.39e-4 & 2.39e-4 & 8.60e-3 & 8.60e-3 & 3.60e+1 & 3.60e+1
\\ \hline
2e+3 & 1.6e-8 & 2.85e-4 & 2.50e-4 & 1.03e-2 & 1.01e-2 & 3.60e+1 &
4.03e+1 \\ \hline
4e+3 & 2.4e-8 & 3.15e-4 & 2.68e-4 & 1.14e-2 & 1.08e-2 & 3.60e+1 &
4.02e+1 \\ \hline
6e+3 & 3.2e-8 & 3.39e-4 & 2.84e-4 & 1.22e-2 & 1.13e-2 & 3.60e+1 &
3.99e+1 \\ \hline
8e+3 & 4.0e-8 & 3.58e-4 & 2.99e-4 & 1.29e-2 & 1.19e-2 & 3.60e+1 &
3.96e+1 \\ \hline
10e+3 & 4.8e-8 & 3.75e-4 & 3.13e-4 & 1.35e-2 & 1.23e-2 & 3.60e+1 &
3.94e+1 \\ \hline
12e+3 & 5.6e-8 & 3.90e-4 & 3.26e-4 & 1.41e-2 & 1.28e-2 & 3.60e+1 &
3.91e+1 \\ \hline
14e+3 & 6.4e-8 & 4.03e-4 & 3.38e-4 & 1.45e-2 & 1.32e-2 & 3.60e+1 &
3.89e+1 \\ \hline
16e+3 & 7.2e-8 & 4.15e-4 & 3.50e-4 & 1.50e-2 & 1.35e-2 & 3.60e+1 &
3.87e+1 \\ \hline
18e+3 & 8.0e-8 & 4.26e-4 & 3.61e-4 & 1.54e-2 & 1.39e-2 & 3.60e+1 &
3.85e+1 \\ \hline
%\label{Table1}
\end{tabular}
\end{center}

\newpage
%\vspace{0.5cm}
\begin{center}
Table 2. Same as Table 1, but the slope at groove root is $m=9.85e-02$.
\vspace{0.5cm}

\begin{tabular}{|c||c||c||c||c||c||c||c|}
\hline
\multicolumn{1}{|c|}{ {step}  } &
\multicolumn{1}{|c|}{ {$t$}  } &
\multicolumn{1}{|c|}{ {$d$,eq.\rf{3.1}} } &
\multicolumn{1}{|c|}{ {$d$,LS M.} } &
\multicolumn{1}{|c|}{ {$w$,eq.\rf{3.2}} } &
\multicolumn{1}{|c|}{ {$w$,LS M.} } &
\multicolumn{1}{|c|}{ {$w/d$,eq.\rf{3.3}} } &
\multicolumn{1}{|c|}{ {$w/d$,LS M.} }\\ \hline \hline
0 & 8.0e-9 & 3.59e-4 & 3.59e-4 & 8.61e-3 & 8.61e-3 & 2.40e+1 & 2.40e+1
\\ \hline
2e+3 & 1.6e-8 & 4.29e-4 & 3.95e-4 & 1.03e-2 & 1.03e-2 & 2.40e+1 &
2.61e+1 \\ \hline
4e+3 & 2.4e-8 & 4.74e-4 & 4.38e-4 & 1.14e-2 & 1.13e-2 & 2.40e+1 &
2.59e+1 \\ \hline
6e+3 & 3.2e-8 & 5.10e-4 & 4.77e-4 & 1.22e-2 & 1.21e-2 & 2.40e+1 &
2.55e+1 \\ \hline
8e+3 & 4.0e-8 & 5.39e-4 & 5.12e-4 & 1.30e-2 & 1.29e-2 & 2.40e+1 &
2.52e+1 \\ \hline
10e+3 & 4.8e-8 & 5.64e-4 & 5.45e-4 & 1.35e-2 & 1.36e-2 & 2.40e+1 &
2.49e+1 \\ \hline
12e+3 & 5.6e-8 & 5.86e-4 & 5.76e-4 & 1.41e-2 & 1.42e-2 & 2.40e+1 &
2.47e+1 \\ \hline
14e+3 & 6.4e-8 & 6.06e-4 & 6.05e-4 & 1.45e-2 & 1.48e-2 & 2.40e+1 &
2.44e+1 \\ \hline
16e+3 & 7.2e-8 & 6.24e-4 & 6.33e-4 & 1.50e-2 & 1.53e-2 & 2.40e+1 &
2.42e+1 \\ \hline
18e+3 & 8.0e-8 & 6.41e-4 & 6.59e-4 & 1.54e-2 & 1.58e-2 & 2.40e+1 &
2.41e+1 \\ \hline
%\label{Table2}
\end{tabular}
\end{center}

\vspace{0.5cm}
\begin{center}
Table 3. Same as Tables 1 and 2, but the slope at groove root is
$m=1.32e-01$.
\vspace{0.5cm}

\begin{tabular}{|c||c||c||c||c||c||c||c|}
\hline
\multicolumn{1}{|c|}{ {step}  } &
\multicolumn{1}{|c|}{ {$t$}  } &
\multicolumn{1}{|c|}{ {$d$,eq.\rf{3.1}} } &
\multicolumn{1}{|c|}{ {$d$,LS M.} } &
\multicolumn{1}{|c|}{ {$w$,eq.\rf{3.2}} } &
\multicolumn{1}{|c|}{ {$w$,LS M.} } &
\multicolumn{1}{|c|}{ {$w/d$,eq.\rf{3.3}} } &
\multicolumn{1}{|c|}{ {$w/d$,LS M.} }\\ \hline \hline
0 & 8.0e-9 & 4.80e-4 & 4.80e-4 & 8.61e-3 & 8.61e-3 & 1.79e+1 & 1.79e+1
\\ \hline
2e+3 & 1.6e-8 & 5.74e-4 & 5.60e-4 & 1.03e-2 & 1.06e-2 & 1.79e+1 &
1.89e+1  \\ \hline
4e+3 & 2.4e-8 & 6.36e-4 & 6.42e-4 & 1.14e-2 & 1.19e-2 & 1.79e+1 &
1.85e+1  \\ \hline
6e+3 & 3.2e-8 & 6.83e-4 & 7.15e-4 & 1.22e-2 & 1.30e-2 & 1.79e+1 &
1.81e+1  \\ \hline
8e+3 & 4.0e-8 & 7.22e-4 & 7.80e-4 & 1.29e-2 & 1.39e-2 & 1.79e+1 &
1.78e+1  \\ \hline
10e+3 & 4.8e-8 & 7.56e-4 & 8.39e-4 & 1.35e-2 & 1.47e-2 & 1.79e+1 &
1.76e+1  \\ \hline
12e+3 & 5.6e-8 & 7.86e-4 & 8.94e-4 & 1.41e-2 & 1.55e-2 & 1.79e+1 &
1.74e+1  \\ \hline
14e+3 & 6.4e-8 & 8.12e-4 & 9.44e-4 & 1.45e-2 & 1.62e-2 & 1.79e+1 &
1.72e+1  \\ \hline
16e+3 & 7.2e-8 & 8.36e-4 & 9.90e-4 & 1.50e-2 & 1.69e-2 & 1.79e+1 &
1.70e+1  \\ \hline
18e+3 & 8.0e-8 & 8.59e-4 & 1.03e-3 & 1.54e-2 & 1.75e-2 & 1.79e+1 &
1.69e+1  \\ \hline
%\label{Table3}
\end{tabular}
\end{center}

\begin{figure}[H]
\centering
\psfig{figure=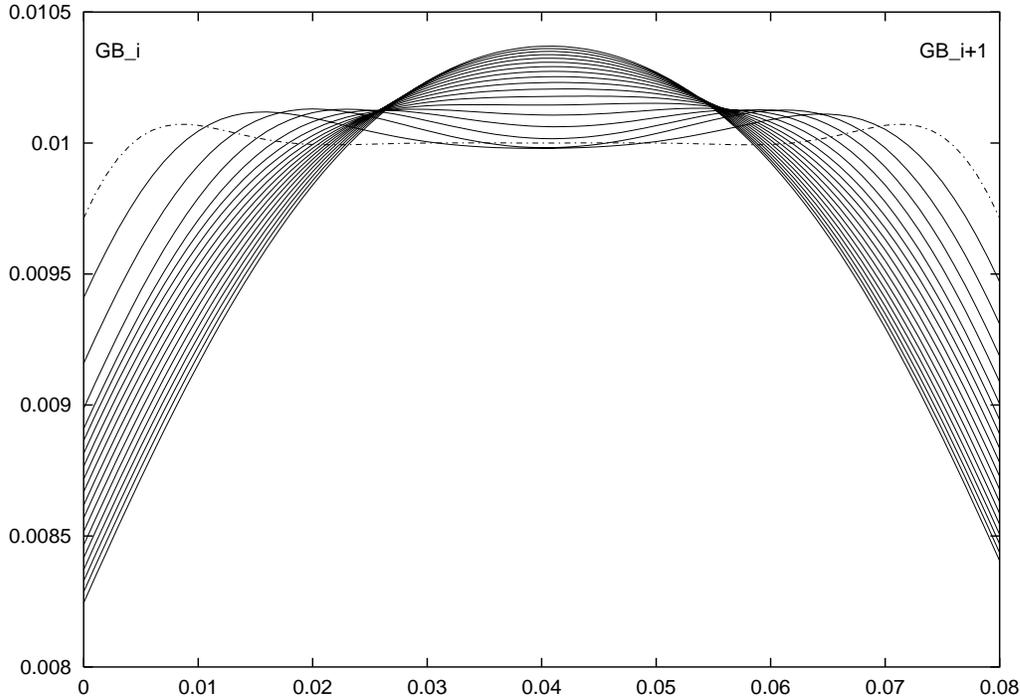,height=3.8in,width=5.5in,angle=270}
\caption{Long-time evolution of surface profile
intersected by two adjacent GBs. The initial surface for LS simulations
is
shown with dashed-dotted line.}
\label{Fig8}
\end{figure}

\bigskip
An interesting simple extension of the classical two-grain model is the
case of a
periodic
array of grains separated by parallel GBs.
In Fig. \ref{Fig8}, we present the results for the evolution of a
surface profile
intersected by two GBs, $i$ and $i+1$. The physical boundary conditions
at
both groove roots are a
constant slope of the surface and zero flux
(for this example, the slope at groove roots is
$m=9.85e-02$). At short times, grooves develop at each grain boundary
according to the solution for an isolated
grain boundary, as presented in Figs. \ref{Fig5} - \ref{Fig7};
grooving stops when, at sufficiently long times, identical
circular arcs develop connecting adjacent GBs. The same result was
obtained in
\cite{HACKNEY} using Fourier method and the SSA.

\bigskip
\Section{Conclusions}
The Level Set method was used to model the grain-boundary grooving
by surface diffusion in an idealized polygranular
interconnect
which consists of grains separated by parallel GBs. The novel feature of
the method
is the treatment of physical boundary conditions at the groove root.
The results obtained are in good agreement with the classical one
(Mullins, \cite{MULLINS57}) for the case of an isolated grain boundary
(two-grain case) and with more recent results of \cite{HACKNEY} for the
case of
periodic array of grains.
One goal  for future work is to apply
electromigration influence on the grooving process.
In addition, the algorithm and its software implementation  will be used
by materials
 scientists to pursue studies of 
%electromigration phenomena with
%different material constants.
GB grooving with an arbitrary electromigration flux, the various 
ratio of the GB to surface diffusivity which was predicted to 
critically affect the groove kinetics and shape  account for 
various EM failure regimes \cite{GN}.

\end{document}